
\newcount\fcount \fcount=0 \def\ref#1{\global\advance\fcount by 1
\global\xdef#1{\relax\the\fcount}} \raggedbottom
\footline={\ifnum\pageno=1 \hfil \else \hss\tenrm\folio\hss\fi}
\def\etal {{\it et~al.}}
\def\la{\lower.5ex\hbox{$\; \buildrel < \over \sim \;$}}
\def\ga{\lower.5ex\hbox{$\; \buildrel > \over \sim \;$}}

\tolerance = 30000
\magnification = 1200
\baselineskip=5.50mm plus 0.1mm minus 0.1mm

\centerline {\bf Radiation Dosimetry of Binary Pulsars}

\vskip 0.5cm

\centerline {David Eichler}

\centerline {Dept. of Physics, Ben Gurion University, BeerSheva, Israel}
\centerline{(eichler@bguvms.bgu.ac.il)}

\centerline {Biman B. Nath}

\centerline{ Inter-University Center for Astronomy \& Astrophysics,}
\centerline{Post Bag 4, Ganeshkhind, Pune - 411007, India}
\centerline{(biman@iucaa.ernet.in)}
\vskip 0.5cm

\centerline{\it Received March 27 1995; \quad accepted July 21 1995}
\vskip 1cm

\centerline{\bf Abstract}
\bigskip
Companion stars exposed to high energy radiation from a primary
neutron star or accreting black hole  can experience  significant spallation
of their heavy elements, so that their atmospheres would be extremely
rich in lithium, beryllium, and especially boron.  In this paper we note
that the detection  or non-detection of these elements, and their
relative abundances if detected, would  provide a diagnostic of the high
energy output of the primary, and possibly the shock acceleration of
particles at the companion's bow shock in  a  pulsar wind.


\vfill

\bigskip\bigskip
\noindent
{\bf 1. Introduction}
\medskip
Pulsar emission theory is by now an established if unresolved topic.
Radio pulsations are believed to imply  counterstreaming pairs, which
are produced via the
curvature radiation of very high energy gamma rays.  Below the ``death
line",  it is believed that the primary radiation is mainly curvature
radiation that fails to develop a next generation of pairs.
Even  while pulsing, a pulsar could easily put out $\sim 10^{-3}$  of its
spin down energy into pulsed gamma rays, (Usov 1983) since a)
pair production shorts out the field to the point that the pairs are
accelerated only enough to marginally produce pairs, and b) charged
particles that are accelerated outward will eventually reach a site where
emitted photons can escape without further pair production.  A reasonable
estimate for curvature radiation is probably the product of the
Goldreich-Julian
current and the polar voltage drop, whence the fraction of total spin down
power  radiated as curvature gamma rays can be of order $10^{-3}$ or
more for millisecond pulsars.
Luminous,  nearby  pulsars such as the Crab are
 observed to yield pulsed gamma rays at a total power consistent with
theoretical expectations. Recent EGRET limits are also consistent
with these expectations (Fierro et. al. 1995).  However,
additional gamma radiation may be generated by pairs striking and cascading
in the atmospheres of companions to pulsars. Also, the pair luminosity
itself may be enhanced by shock acceleration of  pulsar wind
particles at the bow shock of its companion ( Arons and Tavani, 1994,
Grove et. al. 1995).

 In this paper, we  suggest that the output in
high energy ($E>20$ Mev)
quanta from  pulsars with
close binary companions can be diagnosed via the production of
light elements in the companion's atmosphere by photospallation.
The idea of using light elements as a dosimetric diagnostic of
pulsars is not new: it has been discussed in the context of very
young pulsars irradiating early supernova remnants
(Eichler and Letaw 1987).
More recently, Li has been reported to be present
in roughly cosmic abundances in V404 Cygni, contrary to
expectations that it should have been mostly destroyed.
It has been considered that the Li has been replenished
by photospallation caused by irradiation from its compact
(primary) companion. Although this invokes some numerical coincidence,
alternative explanations for the Li abundance in V404 Cygni-like
systems are briefly mentioned below.

The question of whether ``black widow'' type evaporation scenarios occur
for accreting neutron stars and pulsars is still an open one. It has been
suggested that heat induced evaporation can be important for companions
to both accreting neutron stars and pulsars (Eichler and Ko 1988; Ruderman
\etal 1989; Ruderman, Shaham and Tavani 1989) though  the efficiency of mass
loss and extent of ablation have been questioned on both observational and
theoretical grounds (Levinson and Eichler 1991; Eichler 1991; Gedalin and
Eichler 1993). This matter is particularly questionable for low
luminosity pulsars  where some mechanism must be invoked for
converting the spin down power to a form of energy suitable for mass
evaporation.   In the latter case, some scenarios assume that the pulsar
spin down power is somehow converted to soft gamma rays with high
efficiency (Kluzniak \etal 1988; Phinney \etal 1988).  Several eclipsing
pulsars  have been  discovered at this time (Fruchter \etal 1988; Lyne \etal
1990; Johnston \etal 1992) and
the eclipse is suggestive of some mass loss mechanism, though not
necessarily implicative of significant ablation. In the PSR 1957+20
system, for example, viable  eclipse  mechanisms typically require a
plasma frequency at the eclipse site of about 0.1 of the frequency of the
radio waves that are being eclipsed (Gedalin and Eichler 1993).   Assuming
that the outflow proceeds at about  the orbital velocity or somewhat
higher, this implies a mass loss rate of about $10^{13}$g/s.

\bigskip
\noindent
{\bf 2. Photospallation and light elements}
\medskip

The  photospallation cross section for  $C^{12}+\gamma \rightarrow
B^{11} + p$  becomes significant at photon energies above 15 Mev or so.
It averages about 5 millibarns  in the photon energy range 20  to 25 MeV,
or about 1 millibarn per unit logarithm in photon energy at 20 MeV
(Taran and Gorbunov 1967).
Higher energy gamma rays or pairs hitting the companion surface cascade
via bremsstrahlung pair production cycles,  and always pass through this
energy range.   Shock acceleration of pairs followed by synchrotron
emission would convert much of the shock energy to photons of about
several Mev,  so the amount of photospallation predicted by this scenario
depends sensitively on a detailed calculation of  the synchrotron
spectrum.

We consider the consequences of  the hypothesis that some fraction
$\epsilon $  of the pulsar's spin down power L arrives in the form of (or is
converted to) 20 Mev photons. For convenience, assume the average  photon
energy  included  in this fraction is $10^{-5}$ erg.  Then the hard photon
flux  incident on the companion  is
$f = 3 \times 10 ^{14} f_{14.5}$, where
$f_{14.5}=({\epsilon \over
{ 3 \times 10^{-3}}}) L_{35}  D_{11}^{-2}$
photons/cm$^2$ s,   $L_{35}$ is L in
units of $10^{35}$ erg s$^{-1}$,
 and $D_{11}$ is the orbital separation in
units of $10^{11}$ cm.  The lifetime of a carbon nucleus exposed to f is about
$3 \times 10^{12} /f_{14.5}$ seconds.

 Can the top radiation length of material  remain on the surface for more
than $10^{10}$ seconds (in which case the heavy elements would be
completely destroyed)? The mass loss rate for the PSR 1957+20 system is
conservatively estimated to be  of order $10^{12} g$/s, and
could easily be  $10^{13} g$/s (so we define it to be $\dot
M_{13}$ times this amount), from a surface area of $16 \pi R^2_2\times
10^{20} cm^2$, (here $ R_2$ is the radius in units of $2 \times
10^{10}$cm. $R_2 = 1$ corresponds to the companion to PSR 1957+20 filling its
Roche lobe ) or a stripping rate of at least
$2  \times 10^{-9}\dot M_{13}R^{-2}_2$ g/cm$^2$s.  As a radiation length
is  roughly $10^2$ gm/cm$^2$,  the stripping time
could in principle exceed
the lifetime of  heavy nuclei at the surface, if there is no significant
mixing of the surface layers, but it does not appear to
exceed it by such a large factor that the spallation products themselves
would be entirely broken up to helium and hydrogen.    Moreover, the
observed absorption lines, insofar as they affect the continuum, are
consistent  with standard solar abundances (Romani, private
communication). We thus conclude that for PSR 1957+20,
and similar system, the spallation of heavy elements, if it occurs, does not
defeat itself by the total destruction of
either the heavy elements or the products. [Such destruction  could
in principle
be  important for some hypothetical range of parameters, (high $f_{14.5}$,
low $\dot{M}$, requiring suppressed mass evaporation by the
radiation).  In this case however the destruction of
primary heavy elements such as carbon would be the more conspicuous effect.]

Let us first suppose that all of the spallation products that are
produced are eventually evaporated. This implies that the exposure time
at the surface is set by the mass loss rate, independent of
convective mixing.
By the above, the fraction of carbon that is spalled to lighter
nuclei can be
of order  $ 1.5 \times 10^{-2} f_{14.5} {(\dot M / 10^{13}gm/s)}^{-1}R_2^2$,
or unity, whichever is less.  For PSR 1957+20-type
parameters, this is many orders of magnitude
 above the cosmic values of Li, Be and B. The largest
change is in the abundance of boron. The abundance levels depend on the
cross-sections of the reactions that produce and destroy the respective
nuclei, the photon flux and the exposure time. In the case of spallation
that is just below the limit of total destruction of heavy nuclei, the ratios
B$^{10+11}$/H, Be$^9$/H and Li$^7$/H could be as high as
$5.8 \times 10^{-4},\, 5.9 \times 10^{-5},\, 3.7 \times 10^{-6}$
(by number) after spallation (Boyd and Fencl 1991) compared to the cosmic
ratios of $3.4 \times 10^{-10},\, 4.5 \times 10^{-11},\, 2.3 \times
10^{-9}$
respectively (Cameron 1982). We have estimated
these ratios for a photon number spectrum with an index $-2.7$, assuming that
the expected spectrum of 1957+20 will be similar to that of Crab
nebula (Nolan \etal 1993). These ratios change only slightly with the
exact shape of the spectrum  above $10$ MeV. In any case, the spallation
products should be above the otherwise expected levels, if the incident
flux of sufficiently hard gamma  radiation on the companion   is a
small fraction ($10^{-6}$ or more for 1957+20)  of the intercepted spin down
power. We suggest HST observations of the BI resonance lines near $2497
\AA$ which have been used to estimate the boron abundances in halo
stars (Duncan, Lambert and Lembke 1992).
Unfortunately, the reddenning of 1957+20 may make observations in
the UV difficult, although this does not rule out the UV observations of
other sources.
The Li I resonance line at  $6707.8 \AA$ can be
used to measure the lithium abundance (as was done for
V404 Cygni, see below).

The above neglects both dilution due to turbulent convection to deeper
levels, which may occur more rapidly than the stripping time, and
destruction by  subvection to the core.  In discussing this, it is  useful to
note that the  column density of the entire star  is  M/$\pi R^2$,  where M
and R are respectively the mass  and radius of the star, is about $10^9$
radiation lengths. {\it If} the  mass evaporation is such that all of the
layer down  to which the mixing obtains eventually is lost, then in steady
state,  the mixing  has no effect  on the level of spallation products at the
surface.  This includes the case in which the entire star is eventually
evaporated.  If, on the other hand,
 the  mixing depth $\lambda_m$ is larger than the
time integrated ablation depth $\lambda_a$ (which grows linearly in
time),
then the dilution  of spallation products is
simply the ratio of the two $\lambda_m/\lambda_a$.
 While an evaluation of the mixing depth is beyond the scope of this
paper,  we note that for PSR 1957+20, the layer of mass loss can
be conservatively estimated from the dispersion measure of
the eclipsing wind as $10^{12}g/s$ times the age $t$. For
$t$ of order a  spin down time, which for this particular system is
roughly
a Hubble time or more, the lost mass is then about $10^{-2}$ of the
companion mass, so that the astrophysical uncertainty imposed by
our ignorance of turbulent mixing is not all that large. Even if the system
is only $\sim 3 \times 10^7$ yr old, the uncertainty  in
$\lambda_m/\lambda_a$ is several orders of magnitude less
than the maximum light element enhancement.

The light element abundance in the gas that has been blown off from
the companion, is simply the ratio of the rate of accumulation
of  light element nuclei to the stripping rate, apart from the dilution
factor due to convection as discussed in the previous paragraph. The
rate of accumulation of light elements due to spallation has been
calculated by Boyd and Fencl (1991).
In the limit that neither carbon nor lithium
destruction is significant at the surface we can use
their calculations to express the abundances of light elements
 (for the case of $\lambda _m > \lambda _a$) as:

$$\eqalignno{
{B \over H} &\sim \Bigl ({ \lambda _a \over \lambda _m }\Bigr )
8.5 \times 10^{-7} \dot M _{13} ^{-1} R_2^2 f_{14.5} \> , &\cr
{Li \over H} &\sim \Bigl ({ \lambda _a \over \lambda _m } \Bigr )
 10^{-8} \dot M _{13} ^{-1} R_2^2 f_{14.5} \> , &\cr
{Be \over H} &\sim \Bigl ({ \lambda _a \over \lambda _m }\Bigr )
9 \times 10^{-13}\dot M _{13} ^{-1} R_2^2 f_{14.5} \> ,
&(1)\cr}
$$
where, as argued in the preceding paragraph, the ablation depth $\lambda _a$
is a linear function of time and depends on the particular pulsar and its
companion. Eqns(1) show that the resulting B abundance in the companion
can be very important even in the case of deep mixing, much more than those
of Li and Be. We note here that the ratio $\lambda _a / \lambda_m$ for a
companion mass of $0.02$ M$_{\odot}$ can be written as $0.1 \tau_{10}
{\dot M}_{13} R_m^{-1}$, where $R_m$ is the mixing depth in units of $R_2$,
and $\tau_{10}$ is the system age in $10^{10}$yr.

We have considered the possibility that spallation products are destroyed
by nuclear burning.  Probably the companion star  is too cool for this, if it
fills its Roche lobe,   in the case of 1957+20, where the  companion mass
is  only 0.025 and the radius  about 0.3 (in solar units) (Fruchter \etal
1988). The
central temperature  is for these numbers  only about  $1.5  \times
10^{6}$K. In general, a light companion still on the main sequence   could
burn its spallation products; a  star of 0.08 solar masses, for example, has
a central temperature of  $4 \times 10^6$K,  enough for   lithium burning,
which requires only $3 \times 10^6$K (Swenson, Stringfellow and Faulkner
1990), but,
clearly  even a modest amount of bloating would cool the companion  core
too much to  incinerate the spallation products.  The case  against
incineration is even stronger for Be and B, and/or lighter companions,
such as in the case of 1957+20, where the central temperature may  not in
general be enough for incineration under any set of assumptions.

\bigskip
\noindent
{\bf 3. V404 Cygni}
\medskip

The lithium abundance in the secondary in the V404 Cygni system has been
estimated to be of the order of Li/H $\sim 10^{-9}$, which is close to the
cosmic interstellar ratio (Martin {\etal } 1992).
 This is anomalous considering the
fact that the secondary is most probably a G9 dwarf star (Martin \etal 1992),
and a G/K main sequence star should deplete its lithium abundance by an order
of magnitude after $\sim 100$ Myr and by 2-3 orders of magnitude after
$\sim 1$ Gyr. In the light of the above discussion on photospallation, it
seems possible that $\gamma$-radiation from the vicinity of the central
object, a stellar mass black hole, can easily reimburse the lost lithium in
the secondary star, as has been noted by previous authors
(Martin \etal 1992, Gilmore 1992; also see discussions on
photospallation and light elements in other astrophysical
contexts, e.g., Boyd and Fencl 1991, Gnedin and Ostriker 1992) (While
 Martin \etal (1992) proposed particle spallation in the accretion
disk around the primary, spallation at the companion's surface
should work just as well).
 The observed abundance of lithium that is near the
cosmic abundance would require a rough, coincidental balance
between the production and destruction processes (at the present epoch)
described in the eqns (1).
However, there may be two other reasons for the preservation of
Li at  roughly cosmic abundances that do not
require such a coincidence: Firstly, it may be that heating
of the companion surface by the primary compact object merely
stabilizes  the outer layers to convection, so that the light
elements in these layers are never subducted to the
depths at which they are destroyed. Secondly, it may be that
 the black widow effect, if the companion is destroyed on
a timescale of $\sim 10$ Myr,  selects out for observation only
those systems that are too young to have suffered significant
Li destruction.  Theoretical analysis of the possibilities from
first principles, in our view, are extremely difficult, and not in
any case the primary topic of this paper. Their viability depends,
as for the case of pulsar companions,
 on the turbulence induced by horizontal pressure
gradients resulting
from uneven heating of the surface. Such turbulence,
together with heat conduction, competes with outward mass flow in
determining the entropy
distribution in the outer layers of the heated
companion, and the matter is in our view not fully resolved. But if the
two ``preservation" scenarios for the light elements can be
observationally distinguished from the scenario of
``restoration" by  photospallation (the former predicting a higher
ratio of
Li$^7$  to the other light elements), then the results might have
implications for the companions to pulsars as well.

\bigskip
\noindent
{\bf Conclusions}
\medskip

Any model of companion irradiation by a primary
pulsar in which quanta exceeding 15 Mev or so strike the companion
surface  predicts that the companion atmosphere be anomalously rich in
Li, Be and, particularly  B if the central temperature of the
companion is too small to
burn it. Spectroscopic
observations of an ever growing number of companions to  binary pulsars
may prove to be a powerful diagnostic of the high energy output of the
primary pulsar that is predicted theoretically.

\bigskip
We are grateful to Profs. V.V. Usov  and R. Romani
for useful  discussions.  This research
was supported in part by the Israeli Foundation for Basic Research.

\vfill\eject

\bigskip\bigskip\bigskip \centerline{\bf References}\medskip
\def\r{\hangindent=1pc \noindent}

\r{Aron, J, and Tavani, M. 1994, {in Proc. of IAU Colloquium 142},
{\it ApJSupp}, {\bf 90}, 797}

\r{Boyd, R. N. \& Fencl H. S. 1991, {\it Ap.J} {\bf 373}, 84} \par

\r{Cameron, A. G. W.  1982, in {\it Essays in Nuclear Astrophysics}
(ed. Barnes, C. A. et al.) (Cambridge: Cambridge University Press), p. 23} \par

\r{Duncan, D., Lambert, D., Lembke, M. 1992, {\it ApJ} {\bf
401}, 584} \par

\r{Eichler, D., and Letaw, J. R. 1987, {\it Nature}, {\bf 328}, 783} \par

\r{Eichler, D., and Ko, Y.K. 1988, {\it ApJ}, {\bf 328}, 179} \par

\r{Eichler, D. 1991, {\it ApJ Lett.}, {\bf 370}, L27} \par

\r{Eichler, D.  and Levinson, A. 1988, {\it ApJ Lett.},{\bf 335}, L67} \par

\r{Fierro. J. M., et al. 1995, EGRET preprint}

\r{Fruchter, A. S. {\it et al.} 1988, {\it Nature} {\bf 334}, 686} \par

\r{Gedalin, M. and Eichler, D. 1993, {\it ApJ}  {\bf 406}, 629} \par

\r{Gilmore, G. 1992, {\it Nature} {\bf 358}, 108} \par

\r{Gnedin, N. Y. and Ostriker, J. P., {\it ApJ} {\bf 400}, 1} \par

\r{Grove, J. et. al. 1995, {\it ApJ Lett.} (in press)}

\r{Johnston, S. {\it et al.} 1992, {\it ApJ} {\bf 387}, L37} \par


\r{Kluzniak, W., Ruderman, M.,
Shaham, J., Tavani, M. 1988, {\it Nature}, {\bf 334}, 225} \par

\r{Levinson, A. and Eichler, D. 1991, { \it  ApJ} {\bf 379}, 359} \par

\r{Lyne, A. G. {\it et al.} 1990, {\it Nature} {\bf 347}, 650} \par

\r{Martin, E. L., Rebolo, R., Casares, J. \& Charles, P. A. 1992, {\it
Nature} {\bf 358}, 129} \par

\r{Nolan, P. L. {\it et al.} 1993 {\it ApJ} {\bf 409}, 697} \par

\r{Phinney, E.S. {\it et. al.} 1988, {\it Nature}, {\bf 333}, 832} \par

\r{Ruderman, M.A. Shaham, J. Tavani,  M. and Eichler, D. 1989, {\it
ApJ}, {\bf 343}, 292} \par

\r{Ruderman, M.A., Shaham, J., Tavani, M.  1989, {\it ApJ}
{\bf 336}, 507} \par

\r{Swenson, F., Stringfellow, G. S., Faulkner, J. 1990, {\it
ApJ} {\bf 348}, L33} \par

\r{Taran, G. \& Gorbunov, A. 1967, {\it Soviet J. Nucl. Phys.} {\bf 6}, 816}
 \par

\r{Usov, V.V. 1983, {\it Nature}, 305, 409} \par

\end